\documentclass[%
 reprint,
 amsmath,amssymb,
 aps,
prl,
twocollumns
,superscriptaddress]{revtex4-2}

\usepackage{graphicx}
\usepackage{dcolumn}
\usepackage{bm}

\usepackage{amsfonts}
\usepackage{amsmath}
\usepackage{amssymb}
\usepackage{booktabs}
\usepackage{dcolumn}
\usepackage{mathtools}
\usepackage{microtype}
\usepackage{multirow}
\usepackage{natbib}
\usepackage{natmove}
\usepackage{pstricks}
\usepackage{siunitx}
\usepackage{subfigure}
\usepackage{tablefootnote}
\usepackage{txfonts}
\usepackage[normalem]{ulem}
\usepackage{xkeyval}
\usepackage{array}
\usepackage{droidsans}
\usepackage{charter}
\usepackage[T1]{fontenc}
\usepackage{graphicx} 
\usepackage[format=plain,justification=justified,singlelinecheck=false,font={stretch=1.125,small,sf},labelfont=bf,labelsep=space]{caption}
\usepackage{float}
\usepackage{fancyhdr}
\usepackage{fnpos}

\begin{document}

\title{Entangled Interlocked Diamond-like (Diamondiynes) Lattices}

\author{C. M. O. Bastos}
\affiliation{Computational Materials Laboratory, LCCMat, Institute of Physics, University of Bras\'ilia, 70910-900, Bras\'ilia, Federal District, Brazil.}

\affiliation{ International Center of Physics, University of Bras{\'{i}}lia, Bras{\'{i}}lia $70919$-$970$, DF, Brazil}

\author{E. J. A. dos Santos}
\affiliation{Computational Materials Laboratory, LCCMat, Institute of Physics, University of Bras\'ilia, 70910-900, Bras\'ilia, Federal District, Brazil.}

\author{R. A. F. Alves}
\affiliation{Computational Materials Laboratory, LCCMat, Institute of Physics, University of Bras\'ilia, 70910-900, Bras\'ilia, Federal District, Brazil.}

\author{Alexandre C. Dias}
\affiliation{Institute of Physics and International Center of Physics, University of Bras{\'{i}}lia, Bras{\'{i}}lia $70919$-$970$, DF, Brazil}

\author{L. A. R. Junior}
\email{ribeirojr@unb.br}
\affiliation{Computational Materials Laboratory, LCCMat, Institute of Physics, University of Bras\'ilia, 70910-900, Bras\'ilia, Federal District, Brazil.}

\author{D. S. Galvão}
\affiliation{Department of Applied Physics and Center for Computational Engineering and Sciences, State University of Campinas, Campinas, 13083-859, SP, Brazil}

\date{\today}

\begin{abstract}
Diamondynes, a new class of diamond-like carbon allotropes composed of carbon with sp$^2$/sp$^3$-hybridized carbon networks, exhibit unique structural motifs that have not been previously reported in carbon materials. These architectures feature sublattices that are both interlocked and capable of relative movement. Using ab initio simulations, we have conducted an extensive investigation into the structural and electronic properties of five diamondyne structures. Our results show that diamondiynes are thermodynamically stable and exhibit wide electronic band gaps, from 2.2 eV to 4.0 eV. They are flexible yet highly resistant compared to other diamond-like structures. They have relatively small cohesive energy values, consistent with the fact that one diamondyne structure (2f-unsym) has already been experimentally realized. Our results provide new physical insights into diamond-like carbon networks and suggest promising directions for the development of porous, tunable frameworks with potential applications in energy storage and conversion.
\end{abstract}


\maketitle


Carbon is unparalleled in its ability to form a wide array of allotropes, enabled by variations in hybridization states and atomic topology \cite{chen2020topological}. During the past two decades, this structural versatility has led to the discovery of several new allotropes in what is often referred to as the new "golden era" of carbon allotropes \cite{geim2007rise,hirsch2010era}. Notable achievements include graphene \cite{novoselov2004electric}, amorphous monolayer carbon \cite{toh2020synthesis}, $\gamma$-graphyne \cite{li2018synthesis}, graphdiyne \cite{gao2019graphdiyne}, biphenylene network \cite{fan2021biphenylene}, and monolayer fullerene networks \cite{hou2022synthesis}. These materials have redefined the scope of carbon-based systems, combining exceptional mechanical robustness \cite{lee2008measurement}, electronic tunability \cite{zhang2005experimental}, and chemical functionality \cite{dreyer2010chemistry}.

An important family of these experimentally realized new allotropes are the graphyne-like materials, which incorporate sp and sp$^2$/sp$^3$-hybridized carbon networks, from zero \cite{CPLfullereneynes} to three dimensions (3D) \cite{baughman1987structure}. In three dimensions, significant theoretical models have been proposed, including the polyyne-based diamond frameworks introduced by Baughman and Galvão \cite{NatureGalvao} and the family of $n$-diamondynes developed by Costa and collaborators \cite{costa2018n}. These 3D architectures are created by inserting acetylene units between tetrahedral carbon centers, allowing tunable porosity \cite{enyashin2011graphene}, adjustable electronic band gaps \cite{wang2016electronic}, and promising gas adsorption characteristics \cite{wang2015diffusion}, thus laying the groundwork for porous diamond analogs.

Building on this progress, Yang et al. \cite{yang2025diamondiyne} recently reported the first experimental realization of diamondiyne structures \cite{yang2025diamondiyne}. However, despite their structural novelty, key physical properties of these frameworks remain largely unexplored, particularly their internal lattice organization, mechanical response, and electronic characteristics. In this study, we employ \textit{ab initio} simulations to reveal a noteworthy and previously unreported feature in carbon-based materials: the presence of movable interlocked diamond-like lattices, where entangled sublattices maintain a degree of relative mobility.

In this context, we present a comprehensive theoretical investigation of a new family of carbon allotropes. This family contains four 3D carbon diamond-like (diamondiynes) lattices, one of which has already been experimentally realized \cite{yang2025diamondiyne}. Results reveal that these structures possess wide electronic band gaps (up to \SI{4.06}{\electronvolt}), high mechanical stability, and pronounced elastic anisotropy, with Young's moduli ranging from \SIrange{5}{30}{GPa} depending on the crystallographic direction. These findings provide fundamental insights into the physical behavior of diamondiynes and establish a foundation for designing porous, mechanically responsive carbon materials with potential applications in energy storage and nanoscale mechanical systems.


To investigate diamondiyne's structural and electronic properties, we have carried out density functional theory (DFT) simulations \cite{Hohenberg_B864_1964,Kohn_A1133_1965} using the SIESTA code\cite{Soler_2745_2002} with pseudo atomic orbitals (PAOs) \cite{Junquera_235111_2001} as basis sets and the exchange-correlation functional proposed by Perdew-Burke-Ernzerhof (PBE)\cite{Perdew_3865_1996}. For geometry optimizations, electronic and mechanical properties, we have used a double-zeta-polarized (DZP) basis set and a \textbf{k}-mesh grid of $8 \times 8 \times 8$ within the Monkhorst-Pack (MP) scheme to sample the Brillouin zone. Additionally, we included the van der Waals correction DFT-D3 proposed by Grimme \cite{Grimme_1463_2004} and implemented in the SIESTA code\cite{Ehlert_7169_2024}. For the convergence criteria, the residual forces were always smaller than $10^{-3}$~\si{\electronvolt/\angstrom}. To improve the description of the band gap, we performed a single-point calculation using the hybrid functional proposed by Heyd, Scuseria, and Ernzerhof (HSE06)\cite{Heyd_8207_2003,Heyd_124_2006}, adopting the same parameters as in HONPAS\cite{Qin_647_2014}, a SIESTA-based code with the HSE06 hybrid functional implemented. 
%
For the DFT molecular dynamics simulations, we have used a $2 \times 2 \times 2$ supercell with the single-zeta (SZ) basis and a $3 \times 3 \times 3$ \textbf{k}-mesh using the MP scheme, since the large number of atoms significantly increases the computational cost.

\begin{figure}[!t]
    \centering
    \includegraphics[width=1\linewidth]{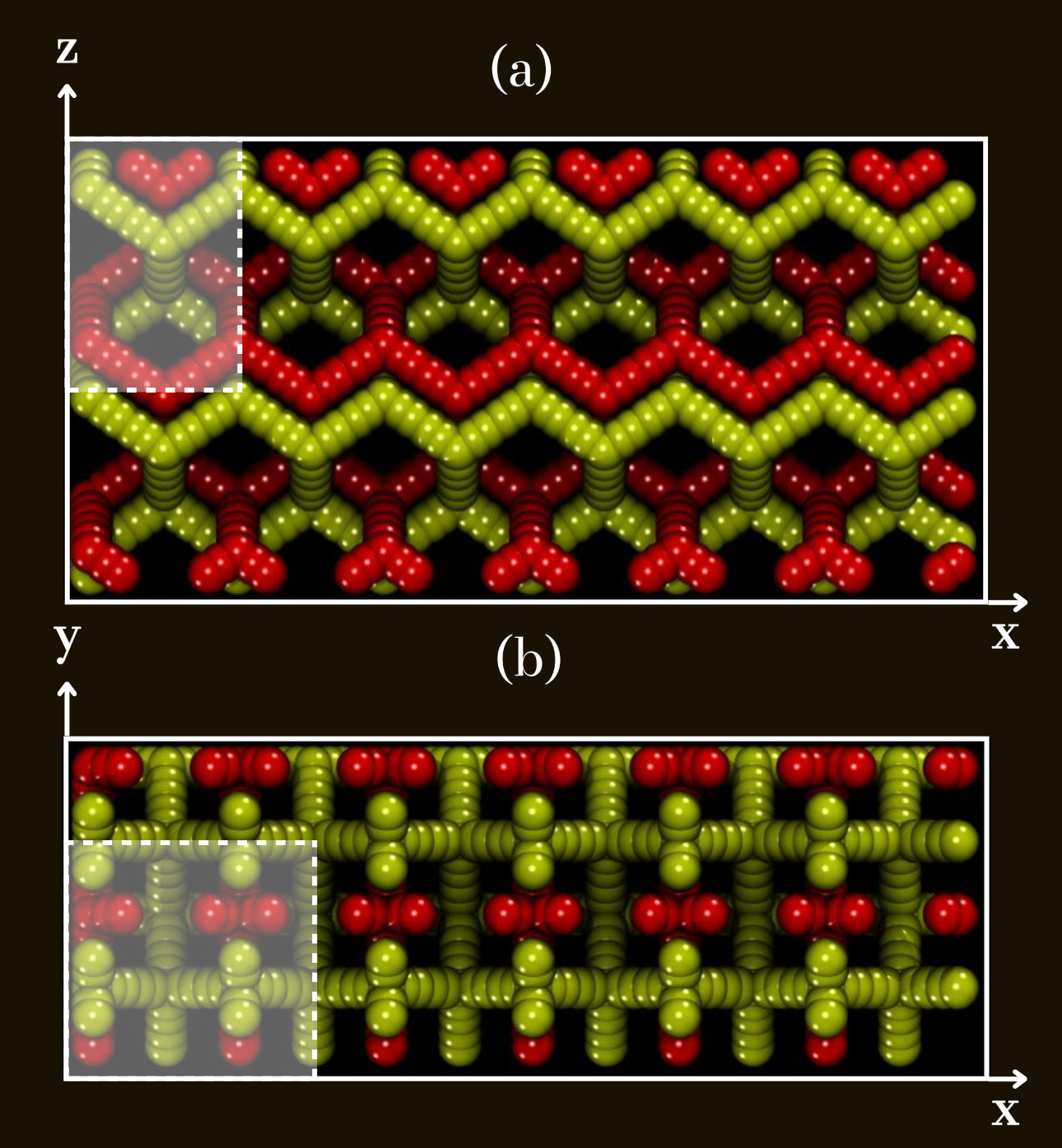}
    \caption{Atomic structure of the 2f-unsym phase, featuring interpenetrated sublattices highlighted in gray and yellow. Each panel displays a projection onto a distinct crystallographic plane: (a) $xz$; and (b) $xy$. The projections show the interlocking between sublattices. The shaded region in each panel indicates the primitive unit cell.}
    \label{fig:2f-unsym-structure}
\end{figure}

The diamondiyne structures reported by Yang et al.\cite{yang2025diamondiyne} are composed of carbon atoms with hybridization $sp$-$sp^3$, where the single and triple bonds are alternated. Among the structural possibilities of the diamondiyne structures, we have selected five structures based on the movable parts with the following characteristics: i) not interpenetrated (ni) with rigid structure; ii) two-folder symmetric (2f-sym) and two-folder unsymmetric (2f-unsym) composed by two movable parts; iii) three-folder (3f) with three movable parts; and iv) four-folder (4f) with four movable parts. The 2f-unsym structure has already been synthesized \cite{yang2025diamondiyne}. This structure is shown in Fig. \ref{fig:2f-unsym-structure}, with projection onto the crystallographic planes $xy$, $yz$, and $zx$. Each movable sublattice is represented in gray and yellow colors, with the highlighted area indicating the unit cell.

Each movable part consists of a sublattice composed of carbon atoms with alternating single and triple bonds. However, there are no covalent bonds connecting the sublattices, which allows significant freedom of movement, restricted only by the entangled region, as illustrated in panel (b) of Fig.~\ref{fig:2f-unsym-structure} for the 2f-unsym structure.
This is the first time that crystalline, interlocked, independent, movable structures have been reported in the literature.

Because of the presence of sublattices, the symmetry of the unit cell is determined by the configuration of the movable parts. In our simulations, we relaxed each unit cell by minimizing both its volume and internal forces. The resulting symmetries were: Fd$\bar{3}$m for \textit{ni}, Pn$\bar{3}$m for \textit{2f-sym}, C2/c for \textit{2f-unsym}, C2/m for \textit{3f}, and P4/nbm for \textit{4f}. The corresponding lattice parameters ($a_0$, $b_0$, and $c_0$ in \AA) are presented in Table~\ref{table:struct}.

Although the lattice parameters vary depending on the symmetry, the number of carbon atoms in the primitive cell remains constant at \num{18}, except for the 2f-unsym structure, which contains \num{36} atoms due to its lower symmetry, requiring a more detailed description of atomic positions. Figures depicting each symmetry and the corresponding SIESTA input files (\texttt{.fdf}) with primitive cells and atomic positions are provided in the Supporting Information.

\begin{table}[!t]
\centering
\caption{Structural parameters for each diamondiyne phase, including space group (SG), number of atoms in the unit cell, lattice parameters ($a_0$, $b_0$, $c_0$ in \si{\angstrom}), cohesive energy per atom $E_{\mathrm{coh}}$ (in \si{\electronvolt}/atom), and distance of average bond (DAV, in \si{\angstrom}). }
\begin{tabular}{lcccccccc} \toprule
Phase & SG & cell & $a_0$ & $b_0$ &$c_0$ & $E_{coh}$ & DAV  \\ 
 &  &    &  (\si{\angstrom}) &(\si{\angstrom})  &(\si{\angstrom})  & (\si{\electronvolt/Atom}) & (\si{\angstrom})\\ \midrule 
NI       & Fd$\bar{3}$m   & \num{18} & \num{11.01} & \num{11.01} & \num{11.01} & -\num{8.82} & \num{1.45} \\
2f-sym   & Pn$\bar{3}$m   & \num{18} & \num{7.85}  & \num{7.85}  & \num{7.85}  & -\num{8.83} & \num{1.35} \\
2f-unsym & C2/c           & \num{36} & \num{11.50} & \num{11.52} & \num{11.52} & -\num{8.89} & \num{1.36}\\
3f       & C2/m           & \num{18} & \num{3.84}  & \num{8.88}  & \num{8.88}  & -\num{8.94} & \num{1.36} \\
4f       & P4/nbm         & \num{18} & \num{6.78}  & \num{6.78}  & \num{4.82}  & -\num{8.97} & \num{1.37} \\
\bottomrule
\end{tabular}
\label{table:struct}
\end{table}

To assess the feasibility of synthesizing the different diamondynes, we have calculated the cohesive energy and found values of approximately $-8.9$~\si{\electronvolt/atom} for all structures. This energy is significantly lower than that of diamond ($\sim -7.4$~\si{\electronvolt/atom})~\cite{mejia2018deorbitalized} and other theoretically predicted diamond-like structures ($\sim -7.1$~\si{\electronvolt/atom})~\cite{shin2014cohesion}, which exhibit only $sp^3$ hybridization. These findings suggest that the alternating single and triple bonds in diamondiynes contribute to their enhanced thermodynamic stability compared to traditional diamond-like structures. The fact that the 2f-unsym phase has already been experimentally realized validates this conclusion.

To further investigate the bonding environment, we have computed the effective coordination number, which was found to be \num{2} for all carbon atoms. This indicates that each carbon has two neighboring carbon atoms, resembling a linear carbon chain. The average bond lengths across all structures are consistent, ranging from \SIrange{1.35}{1.45}{\angstrom}, as summarized in Table~\ref{table:struct}.

\begin{figure}
    \centering
    \includegraphics[width=1\linewidth]{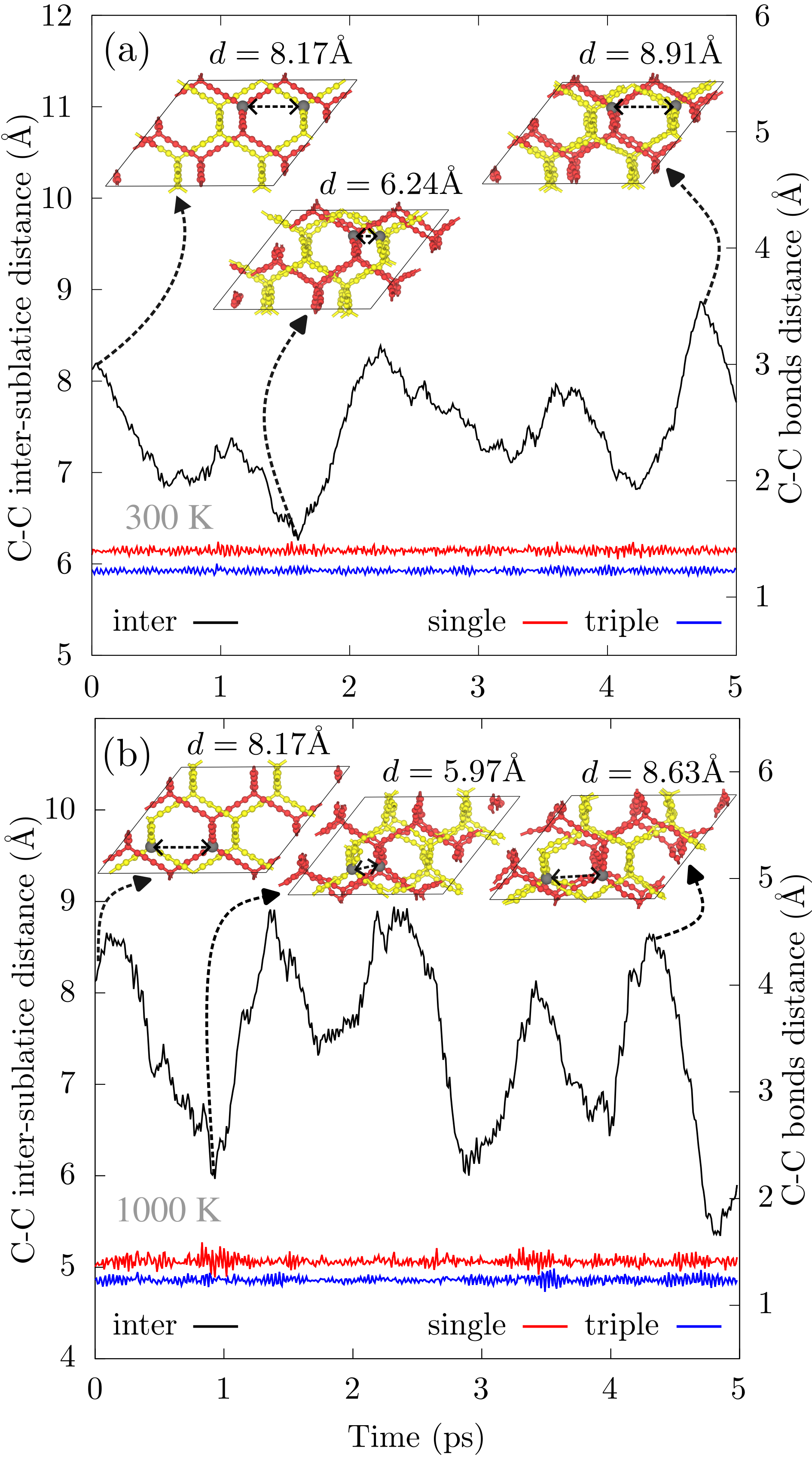}
    \caption{Time evolution of C--C distance values during DFT molecular dynamics simulations. The black line indicates the distance between carbon atoms from different sublattices, with the corresponding scale on the left axis. The red and blue lines indicate the distances between carbon atoms forming single and triple bonds, respectively, within the same sublattice, as referenced on the right axis. Panels (a) and (b) correspond to simulations at \SI{300}{\kelvin} and \SI{1000}{\kelvin}, respectively. Representative structural snapshots illustrate selected time steps, where gray spheres mark the atoms used to monitor the inter-sublattice distances. }
    \label{fig:bonds}
\end{figure}

Figure~\ref{fig:bonds} presents the average bond distances (DAV) for the 2f-unsym structure. We monitor two carbon atoms within the same sublattice, one connected by single bonds (red line) and the other by triple bonds (blue line), and monitor their bond lengths during the DFT molecular dynamics (AIMD) simulations at \SI{300}{\kelvin} and \SI{1000}{\kelvin}, respectively. As expected, the single bonds are longer than the triple bonds. Nevertheless, both remain within the DAV range of approximately \SI{1.35}{\angstrom}, with a variation of about \SI{0.04}{\angstrom}, which is preserved even at \SI{1000}{\kelvin}, despite larger structural fluctuations due to thermal motions.

In contrast, when tracking two carbon atoms from different sublattices, the distance variation is significantly greater, ranging from \SIrange{6.24}{8.91}{\angstrom}. For the 2f-unsym structure, this corresponds to a variation of about \SI{1.93}{\angstrom} at \SI{300}{\kelvin}, nearly \SI{16.7}{\percent} of the unit cell length (approximately \SI{11.5}{\angstrom}). At \SI{1000}{\kelvin}, thermal agitation further increases this variation to \SI{2.66}{\angstrom}, or roughly \SI{23}{\percent} of the unit cell. This behavior is consistently observed for other studied structures, except for the \textit{ni} configuration, which lacks movable parts.

We propose that the high mobility observed in the structure arises from the independent motion of the two sublattices, as illustrated in the snapshot of Fig.~\ref{fig:bonds}, where the gray atoms are used as a fixed reference. At the peaks, these reference points are farther apart, whereas at the valleys, they are closer, as indicated by the dashed line. This motion appears to be constrained by the entanglement between the sublattices.

Throughout our MD simulations, carried out up to \SI{5}{\pico\second}, no periodic pattern is detected in the relative motion, suggesting random and thermally driven dynamics. We attribute this randomness to thermal fluctuations and van der Waals interactions. Furthermore, we observe that the number of displacement peaks increases with temperature, highlighting the competition between thermal agitation and van der Waals forces.

All curves showing the lengths of single and triple bonds, as well as the inter-sublattice distances, are provided in the Supporting Information.


To gain deeper insights into diamondiynes' properties, we have analyzed their electronic properties. The band structures were calculated at $T = 0$\,\si{\kelvin}, using the optimized geometries. The corresponding electronic band gaps, computed using both the PBE and HSE06 hybrid exchange-correlation functionals, are presented in Table~\ref{table:stress}.

As expected, the PBE functional underestimates the electronic band gap values due to the well-known self-interaction error~\cite{bastos2018comprehensive}. Consistent with other diamond-like allotropes reported in the literature~\cite{wang2016electronic}, diamondiynes exhibit wide band gaps ranging from \SIrange{2.208}{4.057}{\electronvolt}, depending on the specific structure.

\begin{figure}
    \centering
    \includegraphics[width=\linewidth]{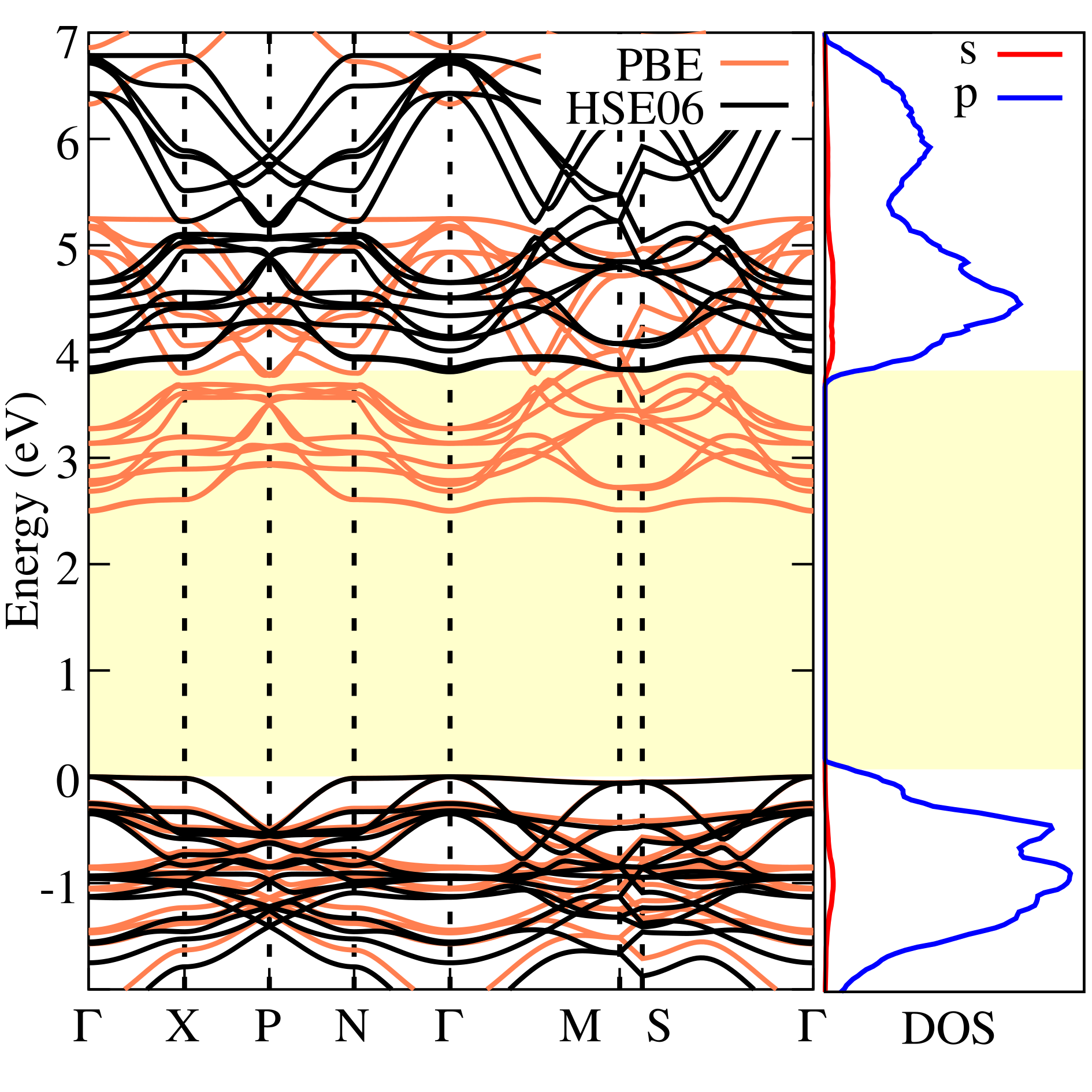}
    \caption{Electronic band structure of the 2f-unsym diamondiyne calculated using the PBE (orange lines) exchange-correlation functional and the HSE06 (black line) hybrid functional. The corresponding density of states (DOS) is also presented, projected onto the $s$ (red line) and $p$ (blue line) orbitals.}
    \label{fig:band-structure}
\end{figure}


We have also evaluated the uniaxial stress-strain behavior. The results show that diamondiynes exhibit high-stress values before failure, ranging from \SIrange{12}{46}{\giga\pascal} depending on the direction and the specific diamondiyne structure. These values are significant compared to most materials, although still below the theoretical strength limit of diamond, which can reach up to \SI{60}{\giga\pascal} \cite{telling2000theoretical}. 

\begin{figure}

    \includegraphics[width=1.0\linewidth]{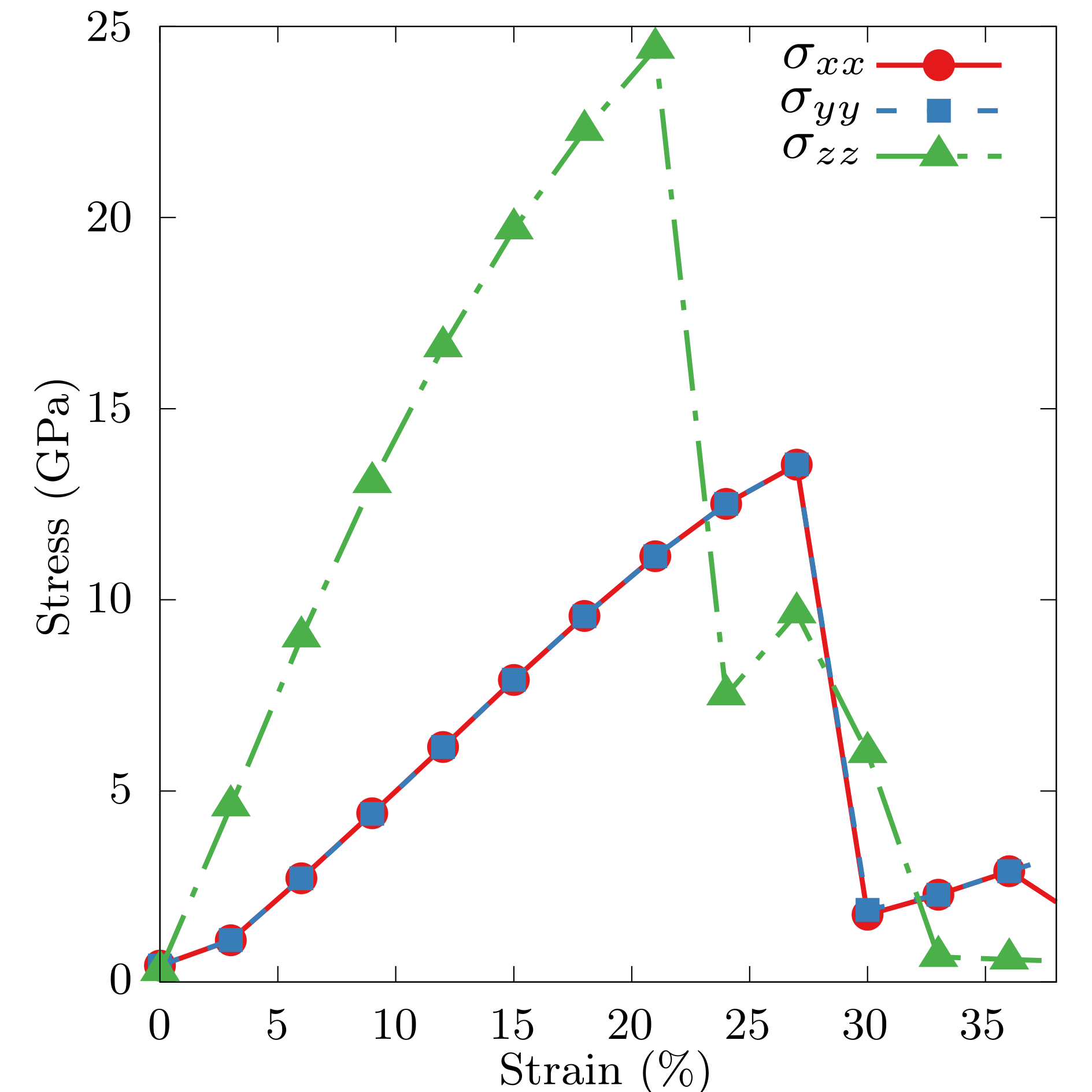}
    \caption{Uniaxial stress-strain curves for the 2f-unsym structure. The uniaxial strain tensor components $\varepsilon_{xx}$, $\varepsilon_{yy}$, and $\varepsilon_{zz}$ were calculated (strain given in \%), and the stress is given in \si{\giga\pascal}.}
    \label{fig:strain-stress}
\end{figure}

In Fig. \ref{fig:strain-stress}, we present the stress-strain curves for the 2f-unsym structure. Due to symmetry, the stresses $\sigma_{xx}$ and $\sigma_{yy}$ are isotropic, whereas $\sigma_{zz}$ displays anisotropy. The strain at failure is approximately \SI{25}{\percent}, comparable to graphene (\SI{25}{\percent}) \cite{varillas2024mechanical} and diamond (\SI{20}{\percent}) \cite{toda2010dft}.

However, unlike diamond, which has only $sp^3$ hybridization, the diamondiynes exhibit mixed $sp$-$sp^3$ hybridization, resulting in a more flexible structure compared to diamond. This combination gives diamondiynes a balance between stiffness and flexibility, making them potential candidates for ultra-resistant and impact-absorbing materials due to their directional stiffness combined with high deformability. 

We have also estimated Young’s modulus values along the $xx$, $yy$, and $zz$ directions. The results are presented in Table \ref{table:stress}. Diamondiynes show a wide range in Young’s modulus values, from \SIrange{19}{137}{\giga\pascal} depending on the structure, symmetry, and direction. Some structures are completely isotropic, such as those with 2f-symmetry, while others are isotropic only in two directions. Within this range, diamondiynes can be considered soft and flexible materials when compared to other carbon allotropes, such as diamond (\SI{1100}{\giga\pascal}) \cite{nie2019approaching} and other theoretically predicted diamond-like (\SIrange{700}{1000}{\giga\pascal}) \cite{de2023nanomechanical}. This exhibit mixed $sp$-$sp^3$ hybridization, whereas diamond has only $sp^3$ hybridization, which results iny $sp^3$ hybridization, which leads to harder materials.

\begin{table}[!t]

\caption{Young's modulus and electronic band gap values for the diamondiynes. The Young's modulus values are presented for the $xx$ ($E_{xx}$), $yy$ ($E_{yy}$), and $zz$ ($E_{zz}$) directions and are given in \si{\giga\pascal}. The electronic band gap values calculated are from PBE and HSE06 hybrid functionals and are given in electron Volts (eV).}

\begin{tabular}{lccccc} \toprule
Phase    &  $E_{xx}$  \quad     & $E_{yy}$  \quad       &$E_{zz}$  \quad      & PBE (gap)          \quad    & HSE06 (gap)   \\ 
         & (\si{\giga\pascal})  & (\si{\giga\pascal})   & (\si{\giga\pascal}) & (\si{\electronvolt})  & (\si{\electronvolt})\\ 
\midrule
NI       & \num{49.2}  & \num{49.2}  & \num{137.3} & 2.999  &    4.057  \\
2f-sym   & \num{36.6}  & \num{36.6}  & \num{36.6}  & 3.000  &    4.051   \\
2f-unsym & \num{19.1}  & \num{19.1}  & \num{75.5}  & 2.499  &    3.809   \\
3f       & \num{56.1}  & \num{56.1}  & \num{62.2}  & 1.062  &    2.208   \\
4f       & \num{49.2}  & \num{49.2}  & \num{137.3} & 1.486  &    2.631   \\
\bottomrule
\end{tabular}
\label{table:stress}
\end{table}

In summary, we have conducted a comprehensive theoretical study of a family of new diamond-like structures, a class of diamondiynes.
Some of the diamondiyne structures present the first reported and unique structural properties of interlocked and movable lattices for diamond-like structures. Diamondiynes with zero to four movable parts were analyzed. All diamondiynes exhibited thermodynamic stability, as confirmed by DFT MD simulations, and have relatively small cohesive energy values, consistent with the fact that one diamondiyne structure (2f-unsym) has already been experimentally realized. They exhibit wide electronic band gap values, ranging from \SIrange{2.2}{4}{\electronvolt}. Additionally, the elastic properties were analyzed, showing that diamondiynes are flexible yet highly resistant compared to other diamond-like structures. Our results provide new physical insights into diamond-like carbon networks and suggest promising directions for the development of porous, tunable frameworks with potential applications in energy storage and conversion.

\bibliography{references.bib}

\end{document}